\documentstyle[epsf,eqsecnum,aps,prb]{revtex} 
\input epsf.tex
\newcommand{\mat}[1]{\mbox{{\boldmath $#1$}}}
\newcommand{\vect}[1]{{\boldmath #1}} 
\newcommand{\onehalf}{{\scriptstyle {1\over  2}}}

\begin{document}
\draft
\twocolumn[\hsize\textwidth\columnwidth\hsize\csname@twocolumnfalse%
\endcsname

\title{Crossover between aperiodic and homogeneous semi-infinite
critical behaviors\\ in multilayered two-dimensional Ising models}

\author{Pierre Emmanuel Berche\cite{byline1} and Bertrand Berche\cite{byline2}}

\address{Laboratoire de Physique des Mat\'eriaux~\cite{byline3}, 
Universit\'e Henri Poincar\'e, Nancy
1, B.P. 239,\\ 
F-54506  Vand\oe uvre les Nancy Cedex, France} 

\date{May, 6 1997, to appear in Phys. Rev. B}

\maketitle

\begin{abstract}
We investigate the surface critical behavior of two-dimensional multilayered aperiodic
Ising models in the extreme anisotropic limit. The system under consideration
is obtained by piling up two types of layers with respectively $p$ and $q$ spin 
rows coupled {\it via} nearest neighbor interactions $\lambda r$ and $\lambda$, 
where
 the succession of layers follows an aperiodic sequence generated through
substitution rules. Far away from the critical regime, the correlation length
$\xi_\perp$, measured in lattice spacing units, in the direction perpendicular to the layers
is smaller than the first layer width and the system exhibits 
the usual behavior of an ordinary surface transition.
In the other limit, in the neighborhood of the critical point, $\xi_\perp$ diverges
and the fluctuations are sensitive to the non-periodic structure of the 
system so that the critical behavior is governed by a new fixed point. We determine the
critical exponent associated to the surface magnetization at the aperiodic 
critical point and show that
the expected crossover between the two regimes is well described by a 
scaling function. From numerical calculations, the parallel correlation length
$\xi_\parallel$ is then found to behave with an anisotropy exponent $z$ which 
depends on the aperiodic modulation and the layer widths.
\end{abstract} 

]

\section{Introduction}
\label{sec:intro}

Since the discovery of quasi-crystals,~\cite{schechtman84} the critical 
properties of aperiodic systems
have been intensively studied (for a review, see 
Ref.~\onlinecite{grimm96}). These systems display, in some cases, the same type
of singularities which can be encountered in random systems, 
 the study of which is a quite active 
 field of research.

In layered systems, generalizing the McCoy-Wu 
model,~\cite{mccoy68a,mccoy68b,mccoy70,auyang74} 
aperiodic distributions of the exchange interactions between successive layers 
in the Ising model have been 
considered.~\cite{igloi88,doria88,benza89,henkel92,lin92,turban93} In 
Ref.~\onlinecite{tracy88}, the Onsager logarithmic singularity of the specific
heat was found to be washed out. The major result was obtained by 
Luck,~\cite{luck93a} by a generalization to layered perturbations of the Harris
criterion for random systems.~\cite{harris74} 

According to Luck's criterion, 
aperiodic modulations may be relevant, marginal or irrelevant, depending
on the correlation length exponent $\nu$ of the unperturbed system, and on
a wandering exponent $\omega$ which characterizes the fluctuations of the couplings
around their average.~\cite{queffelec87,dumont90} Systematic studies of the 
surface critical properties for
irrelevant, marginal, and relevant perturbations have been achieved in the 
extreme anisotropic limit.~\cite{turban94a,turbanigloi94,turban94b,karevski95}
In the case of bulk marginal sequences ({\i.e.} which modifies both surface and
bulk properties), an anisotropic scaling behavior was obtained in the 
Hamiltonian limit,~\cite{berche95,berche96} as well as in the classical 
two-dimensional counterpart,~\cite{igloilajko96} or in hierarchical 
models.~\cite{igloilajkoszalma95} In these systems, both surface and bulk properties
have been studied and
the bulk specific heat exponent, given by $\alpha=1-z$, where $z$ is the anisotropy
exponent, is always negative leading to a non-diverging singular behavior.

In the present paper, we continue the investigation of the surface critical
behavior of marginal sequences, through a study of the crossover between
usual ordinary surface transition and aperiodic critical behavior in 
multilayered
Ising models. Up to now, only layered systems have been considered, and
we study here multilayered systems made of the piling up of two kinds of
layers of widths $p$ and $q$.
In such  systems, 
together with the perpendicular correlation length $\xi_\perp$, the first layer
 width $p$ plays the role
of a 
relevant length scale entering the description of the critical behavior in the vicinity
of the free surface. Depending on the value of the scaled variable $p/\xi_\perp$,
one thus expects a crossover between the two regimes. 
In section~\ref{sec:lay},
we recall some generalities about aperiodic sequences
generated through substitution and  give a short summary of the relevance 
criterion proposed by Luck.
 The critical behavior of the surface 
magnetization in multilayered models at the aperiodic fixed point is 
determined in
Sec.~\ref{sec:surfmag} and the crossover effect, which  constitutes the main
result of the paper is presented in Sec.~\ref{sec:cross}.
In section~\ref{sec:corr}, a numerical study of the parallel correlation 
length $\xi_\parallel$ allows the determination of the anisotropy exponent of
multilayered Ising models in agreement with a relation, already
established in the layered case, which seemingly has a general field of 
application.

\section{Luck's criterion}
\label{sec:lay}

Many results have already been obtained in  $2d$ layered Ising models with 
constant interactions $K_1$ along
the layers and modulated interactions $K_2(k)$ between successive spin rows, $k$
and $k+1$. The transfer matrix should in principle be diagonalized
for arbitrary couplings,~\cite{abraham71} but in the following
one  considers the extreme anisotropic limit. In the Hamiltonian limit,  $K_1\to\infty$, $K_2\to0$, while keeping the ratio
$\lambda_k\!=\! K_2(k)/K_1^*$ fixed, where $K_1^*\!=\!-\frac{1}{2}\ln\tanh
K_1$ is the dual coupling, the row-to-row transfer operator
${\cal T}\!=\!\exp[-2K_1^*{\cal H}]$ involves the Hamiltonian of a quantum
Ising chain in a transverse field:~\cite{kogut79,pfeuty70}
\begin{equation}
{\cal H}=-\frac{1}{2}\sum_{k=1}^L\left(\sigma_k^z+\lambda_k
\sigma_k^x\sigma_{k+1}^x\right)\; , 
\label{eq-1} 
\end{equation}
where the $\sigma$'s are Pauli spin operators. 

The Jordan-Wigner transformation,~\cite{jordan28}  followed by a canonical 
transformation,
 maps Eq.~(\ref{eq-1})
onto a free fermions problem:~\cite{lieb61}
\begin{equation}
{\cal H}=\sum_\alpha\varepsilon_\alpha\left(\eta_\alpha^\dagger
\eta_\alpha-\onehalf\right)\; . 
\label{eq-2}
\end{equation}
The fermion excitations $\varepsilon_\alpha$ follow from the
solution of the linear system
\begin{eqnarray}
\varepsilon_\alpha{\mit\Psi}_\alpha(k)&=&-{\mit\Phi}_\alpha(k)-\lambda_k
{\mit\Phi}_\alpha(k+1)\nonumber\\
\varepsilon_\alpha{\mit\Phi}_\alpha(k)&=&-\lambda_{k-1}{\mit\Psi}_\alpha(k-1)-
{\mit\Psi}_\alpha(k) \label{eq-3}
\end{eqnarray}
which may be rewritten as a single eigenvalue equation, either for
$\vect{\Phi}$ or $\vect{\Psi}$ which are assumed to be normalized and free boundary conditions
$\lambda_0=\lambda_L=0$ are imposed. 

The surface properties are defined as usually by matrix elements: the surface 
magnetization $m_s$ is given by
$\langle 1\vert\sigma_1^x\vert 0\rangle$ 
where $\vert 0\rangle$ is the ground state of $\cal H$ and $\vert 1\rangle$ is the
first excited state. Using the transformation to $\eta$-fermionic operators,  
the surface magnetization 
takes the simple form $m_s={\mit\Phi}_1(1)$. 
The surface energy density $e_s$  involves the
lowest two-fermion excited state $\vert 2\rangle=\eta_1^+\eta_2^+\vert 0\rangle$, and can thus be written
$e_s=\langle 2\vert\sigma_1^z\vert 0\rangle=(\varepsilon_2-
\varepsilon_1 ){\mit\Phi}_1(1){\mit\Phi}_2(1)$.

With an aperiodic modulation of the couplings, one may write:
\begin{equation}
\lambda_k=\lambda r^{f_k}\; ,
\label{eq-4}
\end{equation}
where $f_k$, which may be $0$ or $1$, is determined by the aperiodic
sequence. Let
$n_L=\sum_{k=1}^L f_k$
be the number of modified couplings on a
chain with length $L$; their asymptotic density is
$\rho_\infty\!=\!\lim_{L\to\infty}n_L/L$. The critical coupling $\lambda_c$ 
obeys the following condition:~\cite{pfeuty79} 
\begin{equation}
\lim_{L\rightarrow\infty}\prod_{l=1}^L (\lambda_l)_c^{1/L}=1,
\label{eq-6}
\end{equation}  
which,
using equation~(\ref{eq-4}), gives:
\begin{equation}
\lambda_c=r^{-\rho_\infty}\; .
\label{eq-7}
\end{equation}

The sequences considered below are generated through substitutions rules, 
either on
digits (``period-doubling'') or on pairs of digits (``paper-folding'').  

The ``period-doubling'' (PD)
sequence~\cite{luck93a,turban94a,turban94b,collet80} may be defined by the substitutions ${\cal S}(0)\!=\!1\ 1$ and
${\cal S}(1)\!=\!1\ 0$, which, starting on $1$, give successively:
\begin{equation}  
\begin{array}{lrrrrrrrrr}     
n=1&&1&&&&&&&\\
n=2&&1&0&&&&&&\\
n=3&&1&0&1&1&&&&\\
n=4&\quad&1&0&1&1&1&0&1&0
\end{array}
\label{eq-10}
\end{equation}

The ``paper-folding''
(PF) sequence is generated through substitutions on pairs of
digits, and details can be found in Ref.~\onlinecite{berche96,dekking83a}.

\begin{figure}
\epsfxsize=8.6cm
\begin{center}
\mbox{\epsfbox{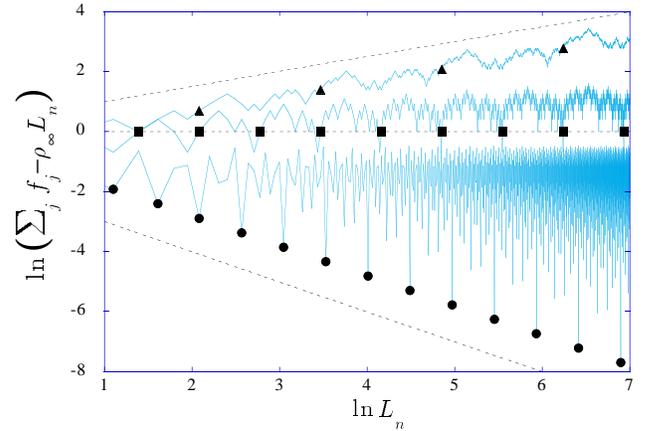}}
\end{center}
\caption{Fluctuations of the sum of the $f_k$'s around their average for three
different aperiodic sequences (top: Rudin-Shapiro, $\omega=1/2$;
middle: ``paper-folding'', $\omega=0$; bottom: Fibonacci, $\omega=-1$). The
symbols correspond to sizes which do respect the symmetries of the chains 
and the dotted
lines have slopes $\omega$.}\label{fluc}  
\end{figure}

Most of the properties of a sequence are  obtained from its substitution
matrix~\cite{queffelec87,dumont90} with entries $M_{ij}$ giving the number
$n_i^{{\cal S}(j)}$ of digits (or pairs) of type $i$ in ${\cal S}(j)$. 
The asymptotic density of modified interactions, $\rho_\infty$, is  deduced from  the  eigenvector corresponding to
the leading eigenvalue $\Lambda_1$ of the substitution
matrix,
allowing the calculation of the critical coupling {\it via}
Eq.~(\ref{eq-7}).  The length of the sequence after $n$ iterations of the 
substitution rules
 is
asymptotically related to the leading eigenvalue 
through  $L_n\sim\Lambda_1^n$. Finally, 
the fluctuations of the $f_k$'s
at a length scale $L_n$ are governed by the next-to-leading eigenvalue 
$\Lambda_2$: 
\begin{equation}
\sum_{k=1}^{L_n}f_k\simeq\rho_\infty L_n+
\vert
\Lambda_2\vert^n F\left({\ln L_n\over\ln\Lambda_1}\right)
\label{eq-12bis}
\end{equation}
where $F(x)$ is a periodic ``fractal function''~\cite{luck93a} as shown in
Fig.~\ref{fluc}.

The cumulated deviation per bond of the couplings from their average 
$\overline{\lambda}$ is thus written
\begin{equation}
\overline{\delta (L_n)}={1\over L_n}\sum_{k=1}^{L_n}(\lambda_k-
\overline{\lambda})\sim\delta\vert
\Lambda_2\vert^n\sim\delta L_n^\omega
\label{eq-12}
\end{equation}
where $\delta\!=\!\lambda(r-1)$ is the amplitude of the modulation and $\omega$ 
is the wandering exponent given by:
\begin{equation}
\omega={\ln\vert\Lambda_2\vert\over\ln\Lambda_1}. 
\label{eq-13}
\end{equation}
Since the relevant length scale is the correlation length $\xi$, 
the aperiodicity thus induces a thermal perturbation
\begin{equation}
{\overline{\delta (\xi)}\over t}\sim t^{-\phi}\; ,\qquad\phi=1+\nu(\omega-1)\; 
\label{eq-14}
\end{equation}
involving a crossover exponent $\phi$.~\cite{luck93a}
When $\phi>0$, the perturbation diverges as $t\to 0$, corresponding to a
 relevant 
perturbation.  When $\phi<0$, the perturbation  is washed out when the
critical point is approached; it is irrelevant. When $\phi\!=\!0$, the perturbation
is marginal and may lead to a non-universal behaviour with 
perturbation-dependent 
critical exponents. In the following, we will only consider marginal sequences.

In the case of the $2d$ Ising model, since $\nu=1$, the relevance of the 
perturbation
is directly determined by $\omega$ as shown in Fig.~\ref{fluc}.

\section{Surface magnetization in multilayered Ising models}
\label{sec:surfmag}

\subsection{Critical point and surface magnetization}

 The first excited state  of the Hamiltonian, $\vert 1\rangle$,
is degenerate with the ground state in the ordered phase of the infinite
system
and, since a simplification occurs in Eq.~(\ref{eq-3})  when $\varepsilon_1$ 
vanishes, the first relation
 provides a recursion for the components of $\vect{\Phi}_1$. 
After normalization one obtains the surface magnetization of the
semi-infinite system as:~\cite{peschel84}   
\begin{equation} 
m_s=S^{-1/2},\quad S=1+\sum_{j=1}^\infty\prod_{k=1}^j\lambda_k^{-2}
\label{eq-15}
\end{equation}
which remains valid for any distribution of the interactions.

We consider here a multilayered Ising model where the aperiodic sequence 
$f_k=1,\ 0,\ 1,\ 1\dots$
refers now to successive layers with $p,\ q,\ p,\
p\dots$ spin rows. A layer labelled by index $k$ contains $p$ (respectively $q$) 
exchange interactions $\lambda r$ (respectively $\lambda$) between nearest 
neighbor rows according to the
following scheme showing the first three layers:
\begin{equation}\matrix{&
\overbrace{\lambda r\quad \lambda r\quad \dots\lambda r}^{p\ \hbox{{
\rm\scriptsize terms}}}&
\overbrace{\lambda \quad \lambda \quad \dots\lambda}^{q\ \hbox{{\rm\scriptsize terms}}}&
\overbrace{\lambda r\quad \lambda r\quad \dots\lambda r}^{p\ \hbox{{\rm\scriptsize terms}}}
\cr
k&1&2&3\cr
f_k&1&0&1\cr
l&1\ 2\dots p&p+1\dots p+q&p+q+1\dots 2p+q\cr
}
\end{equation}

Let us first determine the critical coupling $\lambda_c$. The width of
layer $k$ is written
\begin{equation}
m_k =(p-q)f_k +q.
\label{eq-8}
\end{equation}  
On the system containing $L$ layers, one has $n_L$ layers
of type $p$ and $L-n_L$ of type $q$ and the critical value is still determined through
Eq.~(\ref{eq-6})
where 
\begin{equation}
\prod_{l=1}^L (\lambda_l)_c^{1/L}=(\lambda_c
r)^{pn_L/L}(\lambda_c)^{q(L-n_L)/L}.
\label{eq-9}
\end{equation}  
 In the thermodynamic limit, it yields
\begin{equation}
\lambda_c =r^{-{p\rho_\infty\over (p-q)\rho_\infty +q}}\ .
\label{eq-20}
\end{equation}  

The aperiodic series (\ref{eq-15}) entering the surface magnetization has the form 
\begin{eqnarray}
S^{(m)}(\lambda&,&r)=1+\sum\limits_{l=1}^{m_1}(\lambda r^{f_1})^{-2l}+(\lambda
r^{f_1})^{-2m_1}\sum\limits_{l=1}^{m_2}(\lambda r^{f_2})^{-2l}\nonumber\\
&&+(\lambda r^{f_1})^{-2m_1}(\lambda
r^{f_2})^{-2m_2}\sum\limits_{l=1}^{m_3}(\lambda r^{f_3})^{-2l}+\dots
\label{eq-21}
\end{eqnarray} 
where the superscript $(m)$ is  for the multilayered system. 
This expression may be shortened with the notations 
\begin{equation}
\sigma_k =\sum_{l=1}^{m_k}(\lambda
r^{f_k})^{-2l}=\left\{\matrix{{1-(\lambda r)^{-2p}\over (\lambda
r)^2-1}=a_p\ &\hbox{\rm if}\  &f_k  =1\hfill\cr\noalign{\vskip 2mm} 
{1-\lambda^{-2q}\over\lambda^2
-1}=b_q\ &\hbox{\rm if}\  &f_k =0\hfill\cr}\right.
\label{eq-22}
\end{equation} 
and $g_k =(\lambda r^{f_k})^{-2m_k}$. Equation~(\ref{eq-21}) then becomes
\begin{equation}
S^{(m)}(\lambda,r)=1+b_q \Sigma(\lambda,r) +(a_p -b_q)\Sigma'(\lambda,r)
\label{eq-23}
\end{equation} 
where
\begin{eqnarray}
\Sigma(\lambda,r)&=&\sum_{j=0}^\infty\prod_{k=0}^j g_k,\nonumber\\
\Sigma'(\lambda,r)&=&\sum_{j=0}^\infty f_{j+1}\prod_{k=0}^j g_k,
\label{eq-24}
\end{eqnarray}
and $f_0\equiv 0$. 
 In the second sum, $\Sigma'$, $f_{j+1}$ may be rewritten as
 \begin{equation}
f_{j+1}={(\lambda^{p-q}r^p)^{-2f_{j+1}}-1\over (\lambda^{p-q}r^p)^{-2}-1}
\label{eq-25}
\end{equation}
and Eq.~(\ref{eq-23})  translates into
\begin{eqnarray}
S^{(m)}(\lambda,r)=1&-&{\lambda^{2q}(a_p -b_q)\over 
(\lambda^{p-q}r^p)^{-2}-1}\nonumber\\
&+&\left[b_q+{(\lambda^{2q}-1)(a_p-b_q)\over (\lambda^{p-q}r^p)^{-2}-1}\right
]\Sigma(\lambda,r).
\label{eq-26}
\end{eqnarray}

The critical behavior of the surface magnetization is governed by the sum 
$\Sigma$, for which an identity can be found, relating it to the aperiodic
series $S^{(l)}$ of the usual layered system in the case $p=q=1$. 
Taking account of Eq.~(\ref{eq-4}),
the series $S^{(l)}(\lambda ,r)$ in  Eq.~(\ref{eq-15})
takes the form
\begin{equation}
S^{(l)}(\lambda,r)=\sum_{j=0}^\infty\lambda^{-2j}r^{-2n_j}
\label{eq-32}
\end{equation}
where the upperscript $(l)$ stands now for the layered problem. One usually
studies this expression for specific sequences,~\cite{turban94a,karevski95,berche96} 
but for our purpose a more
convenient expression can be  
 rewritten as a function of the deviation from the critical 
point:\cite{notet}
\begin{equation}
t=1-\left(\frac{\lambda_c}{\lambda}\right)^2.
\label{eq-33}
\end{equation}
Equation~(\ref{eq-32}) becomes
\begin{equation}
\widetilde S^{(l)}(t,\lambda_c)=\sum_{j=0}^\infty (1-t)^j
\lambda_c^{-2(j-n_j/\rho_\infty)}.
\label{eq-34}
\end{equation}

The 
analysis of
the sum 
$\Sigma$ is now facilitated, since the following identity holds:
\begin{equation}
\Sigma(\lambda,r)=S^{(l)}(\lambda^q,\lambda^{p-q}r^p).
\label{eq-39}
\end{equation}
With the variable $t$, it becomes
\begin{equation}
\widetilde\Sigma(t,\lambda_c)=\sum_{j=0}^\infty (1-t)^{(p-q)n_j+qj} 
(\lambda_c^q)^{-2(j-n_j/\rho_\infty)}
\label{eq-40}
\end{equation}
and the critical point behavior reduces to the same series as the layered 
problem up to the transformation $\lambda_c\to\lambda_c^q$:
\begin{equation}
\widetilde\Sigma(0,\lambda_c)=\sum_{j=0}^\infty 
(\lambda_c^q)^{-2(j-n_j/\rho_\infty)}=\widetilde S^{(l)}(0,\lambda_c^q).
\label{eq-47}
\end{equation}
This expression is the basis of further asymptotic analyses using the scaling 
method proposed by Igl\'oi.~\cite{igloi86}
Let  a power series $M(x)$ have a power law singularity at the transition point
$x=x_0$: $M(x)\sim (1-x/x_0)^\beta$. The truncated series of the first $L$
terms at the critical point thus have the following behavior 
$M_L(x_0)\sim L^{-\beta}$ which allows the determination of $\beta$.
Equation~(\ref{eq-47}) also holds for the truncated series and
leads to a simple relation between the critical exponents of the layered 
problem and the multilayered one:
\begin{equation}
\beta_s^{(m)}(\lambda_c)=\beta_s^{(l)}(\lambda_c^q).
\label{eq-43}
\end{equation}

\subsection{Period-doubling and paper-folding sequences}

\begin{figure}
\epsfxsize=8.6cm
\begin{center}
\mbox{\epsfbox{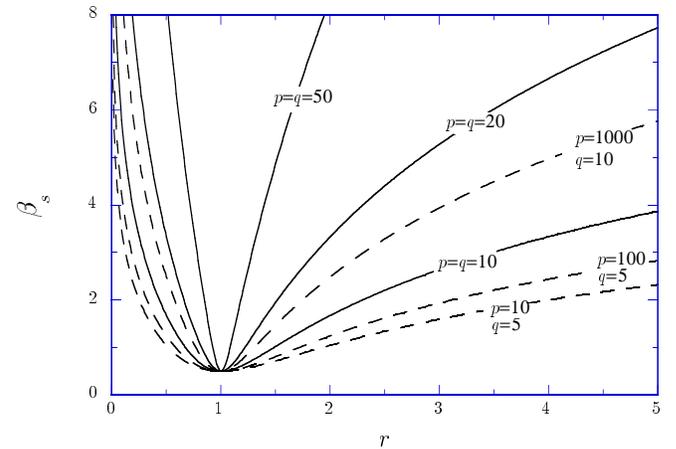}}
\end{center}
\caption{Marginal variation of the surface magnetization exponent $\beta_s^{(m)}$
with the coupling ratio $r$ for the period-doubling multilayer.
The singularity is always weaker than in the
homogeneous system and the surface transition is of second order for
any value of $r,p,q$.}\label{betas-pd}  
\end{figure}

The period-doubling sequence, given after several substitutions in 
Eq.~(\ref{eq-10}), corresponds to a vanishing wandering exponent and an 
asymptotic
density of modified couplings  $\rho_\infty =2/3$, leading to the critical  
coupling for
the  multilayered Ising model
coupling {\it via} Eq.~(\ref{eq-20}):
\begin{equation}
\lambda_c=r^{-\frac{2p}{2p+q}}.
\label{eq-42}
\end{equation}
Equation~(\ref{eq-40}) can be rewritten as
an infinite product,\cite{turban94a}
 whose first $l$ terms contain, at the critical point $t=0$, the first 
 $L=2^{2l}$ terms of the
sum~(\ref{eq-34}) which becomes
\begin{equation}
\widetilde\Sigma_{L=2^{2l}}(0,\lambda_c)=[(1+\lambda_c^q )(1+\lambda_c^{-q})]^l \sim
(2^{2l})^{2\beta_s^{(m)}}.\label{eq-30}\end{equation}
The surface exponent then follows:\cite{igloi86}
\begin{equation}
\beta_s^{(m)} (\lambda_c)={\ln [(1+\lambda_c^q )(1+\lambda_c^{-q})]\over 4\ln 2}.\label{eq-31}
\end{equation}
Going back to the original parameters, one finally obtains:
\begin{equation}
\beta_s^{(m)}(r,p,q)=\frac{\ln(r^{\frac{pq}{2p+q}}+r^{-\frac{pq}{2p+q}})}{2\ln 2}.
\label{eq-45}
\end{equation}
The surface magnetization exponent depends on the amplitude of the
coupling ratio $r$ through $\lambda_c$ as expected for a marginal sequence, but also
on the layer widths. 
The variation of $\beta_s^{(m)}$ for several values of the parameters
is shown in Fig.~\ref{betas-pd}.

The analysis of the series (\ref{eq-40})  can be 
completed by a usual finite-size-scaling 
argument,\cite{barber83} which implies that the following critical point
behavior also holds: $\widetilde\Sigma_L(0,\lambda_c)\sim L^{2x_{m_s}}$. 
The anomalous scaling dimension of the surface
magnetization  $x_{m_s}=\beta_s/\nu$ then takes the same value than $\beta_s$, 
and
this requirement imposes
 that
the correlation length exponent   in the perturbed system keeps
its unperturbed value 
$\nu=1$. This property holds for any marginal sequence and can be understood
within Luck's criterion, which has to remain valid in the perturbed system: the
wandering exponent being constant, $\phi=0$ indeed implies the constancy of 
$\nu$.

In the case of the paper-folding sequence with $\rho_\infty=1/2$, the critical 
point is determined by 
\begin{equation}
\lambda_c=r^{-\frac{p}{p+q}}.
\label{eq-46}
\end{equation}

\begin{figure}
\epsfxsize=8.6cm
\begin{center}
\mbox{\epsfbox{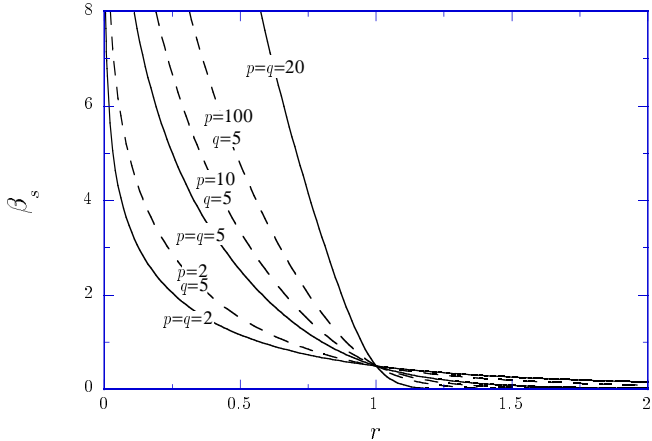}}
\end{center}
\caption{Marginal variation of the surface magnetization exponent $\beta_s^{(m)}$
with the coupling ratio $r$ for the paper-folding multilayer. The singularity
is weakened by the perturbation when $r<1$.
}\label{betas-pf}  
\end{figure}

The same scaling analysis has been done in the case of the paper-folding 
sequence for the layered problem.\cite{berche96} Translated to the
multilayered problem, it leads to
\begin{equation}
\widetilde \Sigma_{L=2^{l}}(0,\lambda_c)=(1+\lambda_c^{2q})^l\sim
(2^{l})^{2\beta_s}
\label{eq-37}
\end{equation}
and to the surface exponent:
\begin{equation}
\beta_s (\lambda_c)={\ln (1+\lambda_c^{2q})\over 2\ln 2}=
\frac{\ln\left( 1+r^{-\frac{2pq}{p+q}}\right)}{2\ln 2},\label{eq-38}
\end{equation}
whose variations are shown  in Fig.~\ref{betas-pf}. While period-doubling is a symmetric sequence
where the relation $\beta_s^{(m)}(r,p,q)=\beta_s^{(m)}(r^{-1},p,q)$ holds, it
is not the case of the paper-folding perturbation for which the surface magnetic exponent
has a monotonous variation. The singularity is thus weakened for $r<1$ only. This 
result can be understood if one looks to  the density of perturbed layers 
(of type $p$). Their density close to the surface is indeed larger than the 
asymptotic density 
$\rho_\infty$, leading to a weaker average coupling in the vicinity 
of the surface than in the bulk when $r<1$. In the other regime $r>1$, the average
coupling close to the surface is enhanced and a stronger singularity follows.

\section{Crossover  in multilayered Ising models}
\label{sec:cross}

The study of aperiodic multilayers does not present more difficulties than
the usual layered systems, but it can produce an interesting crossover
in the critical behavior. While the perpendicular correlation length $\xi_\perp$ is of the order of the size of
the chain in the close neighborhood of the critical point, in which case the critical properties
are governed by the aperiodic fixed point, this is no longer true when the system is
moved far away from $\lambda_c$. In this latter situation, the correlation length
decreases as $\lambda$ increases and, at some point, becomes of the order 
of the first layer width $p$. The aperiodic structure of the system then becomes irrelevant
and the behavior is controlled by the semi-infinite homogeneous fixed point,
 the unperturbed exponent $\beta_s=1/2$ of the ordinary surface transition
  being recovered.~\cite{binder83}

\begin{figure}
\epsfxsize=8.6cm
\begin{center}
\mbox{\epsfbox{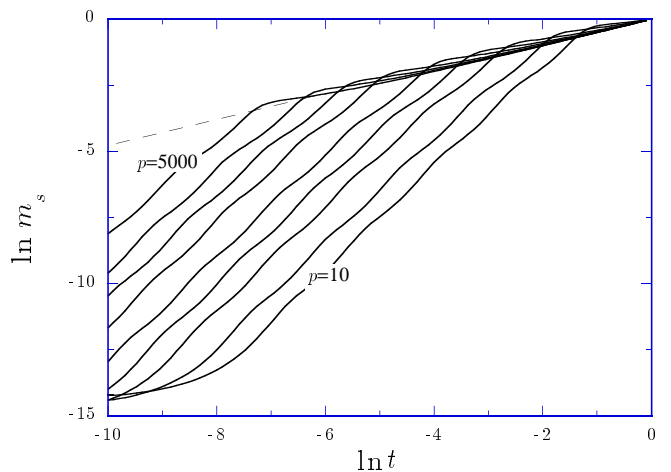}}
\end{center}
\caption{Crossover between homogeneous fixed point ($\beta_s=1/2$) towards
aperiodic fixed point (here $\beta_s^{(m)}=2$) when the critical point is 
approached (period-doubling sequence). The layer sizes are $p=q=10$ ($r=2.295$), $p=20$ ($r=1.515$),
$p=50$ ($r=1.181$), $p=100$ ($r=1.087$), $p=200$ ($r=1.042$), $p=500$
($r=1.017$), $p=1000$ ($r=1.008$), $p=2000$ ($r=1.004$) and $p=5000$
($r=1.002$).}\label{log-log-ms}  
\end{figure}

The evidence
of this
phenomenon 
appears immediatly on a log-log plot as shown in Fig.~\ref{log-log-ms} in
the case of period-doubling sequence. The
values of $r$ and $p=q$ have been chosen for clearness in order to keep a 
constant value
 $\beta_s^{(m)}=2$.
 
\begin{figure}
\epsfxsize=8.6cm
\begin{center}
\mbox{\epsfbox{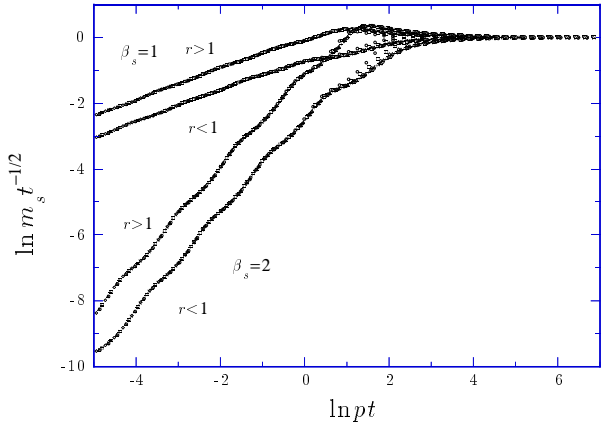}}
\end{center}
\caption{Universal curves for the surface magnetization in the
period-doubling sequence with $\beta_s^{(m)}=2$ and $1$ and $p=q=10$, $20$,
$50$, $100$, $200$, $500$, $1000$, $2000$, $5000$.}\label{log-log-scal}  
\end{figure}

 The surface magnetization can thus be written as follows:
\begin{equation}
m_s(t,p)=t^{1/2}f(p/\xi_\perp)
\label{eq-49}
\end{equation}
where $f(x)$ is a scaling function involving the relevant length scales which
describe the surface behavior. It is shown on a log-log scale in 
Fig.~\ref{log-log-scal} where
$m_st^{-1/2}$ is plotted against $pt$ and $\xi_\perp\sim t^{-1}$.

\begin{figure}
\epsfxsize=8.6cm
\begin{center}
\mbox{\epsfbox{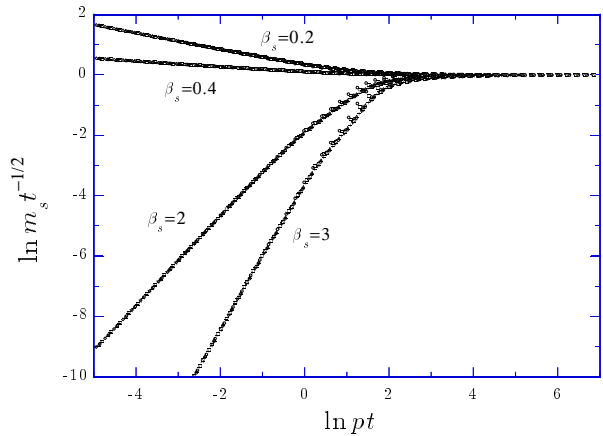}}
\end{center}
\caption{Universal curves for the surface magnetization in the
paper-folding sequence with $\beta_s^{(m)}=3,2,0.4,0.2$  and $p=q=10$, $20$,
$50$, $100$, $200$, $500$, $1000$, $2000$, $5000$.}\label{log-log-scal-pf}  
\end{figure}

For a fixed value of $\beta_s^{(m)}$, one obtains two different universal
curves, one for $r>1$ and the other for $r<1$. This is due to the properties
of the period-doubling sequence which allows to reach the same $\beta_s^{(m)}>1/2$
with two different values of the coupling ratios $r$ at fixed $p$ and $q$ values.
The paper-folding sequence exhibits a rather different behavior shown in 
Fig.~\ref{log-log-scal-pf}.

The scaling function of Eq.~(\ref{eq-49}) has to satisfy appropriate
asymptotic behaviors in order to provide a correct description of the two
limiting power-law behaviors of $m_s$. Far away from $t=0$ it has to keep
a constant value (the usual surface magnetization amplitude of the homogeneous
system) to agree with $m_s\sim t^{1/2}$. This requirement is satisfied
in Fig.~\ref{log-log-scal} and~\ref{log-log-scal-pf} and is responsible for 
the vanishing
slopes for large values of $pt$. On the other hand, in order to recover
the right varying exponent $\beta_s^{(m)}$ close to $t=0$, the following power-law is
expected in the vicinity of the critical point
\begin{equation}
f(x)\sim x^\theta,\ x\to 0,\ \theta=\frac{\beta_s^{(m)}-1/2}{\nu}.
\label{eq-50}
\end{equation}
The average slopes calculated numerically in this second regime are 
in agreement with the expression of $\theta$ given above.

The  scaling functions shown in Figs.~\ref{log-log-scal} and
\ref{log-log-scal-pf} exhibit a quite good agreement with the required asymptotic 
behaviors. In the homogeneous surface regime (far away from $t=0$), 
this fact should be surprising since $t$ is the deviation from the 
{\it  aperiodic
critical point}, given in Eq.~(\ref{eq-33}). Introducing the homogeneous critical
coupling $\lambda_{c,0}=r^{-1}$ of the first (perturbed) layer, and the 
corresponding deviation $t_0=1-(\lambda r)^{-2}$, one has $t\simeq 
t_0(1+\delta/t_0)$, 
where $\delta=1-r^2\lambda_c^2$ is a measure of the difference between the two
critical point locations. Since $\delta$ is a small quantity in our calculations 
(for example
$\delta\simeq 0.005$ in the case $p=q=1000$, $r=1.008$ in 
Fig.~\ref{log-log-scal}), we can expand Eq.~(\ref{eq-49}) up to linear order in 
$\delta$:
\begin{equation}
m_s(t,p)\sim t_0^{1/2}\left(1+\frac{\delta}{2t_0}\right)f
\left(\frac{p}{\xi_\perp}\right),
\label{eq-66}
\end{equation}
and there appears simply a correction to scaling term which is proportional to the
difference between the two critical couplings. With the range of values used here,
this correction to scaling term is negligible.

We can also point out log-periodic oscillations exhibited by the scaling function
in the ``aperiodic'' regime.\cite{karevski96} This behavior is related to the discrete 
scale invariance of aperiodic sequences. The oscillating amplitude,  
 $A$, is  usually a periodic function of $\ln\xi_\perp /\ln m$ where $m$ is the discrete
 scale factor ($4$ for period-doubling and $2$ for paper-folding). In this multilayered system,
the correlation length is naturally measured in units of $p$ such that the
scaling function in the neighborhood of the critical point can be rewritten
\begin{equation}
f(x)\sim A(\ln x/\ln m)x^\theta,
\label{eq-51}
\end{equation}
leading to 
\begin{equation}
\ln f(x)\sim \ln [A(\ln x/\ln m)] +\theta\ln x.
\label{eq-59}
\end{equation}
in agreement with the oscillations around an average slope $\theta$.

\section{Correlation length and marginal anisotropy}
\label{sec:corr}

In Ref.~\onlinecite{berche96}, the low-energy excitation behavior has 
been studied for the layered problem,
leading to a finite-size power law at the critical point involving an 
anisotropy exponent $z$. The conjecture 
\begin{equation}
z=x_{m_s}+\bar x_{m_s},
\label{eq-65}
\end{equation}
was   proposed,\cite{z} and it  has then been proved for specific sequences,
using renormalization group calculations.\cite{igloi96} and has furthermore 
been supported for an arbitrary marginal perturbation
by an approximate calculation of $\varepsilon_1(L)$,~\cite{igtukasz97} 
leading to
$\varepsilon_1(L)\sim m_s(L)\bar m_s(L)\prod_{i=1}^{L-1}\lambda_i^{-1}$ 
from which 
Eq.~(\ref{eq-65})
follows. Moreover, the constancy of the correlation exponent in the perturbed
systems has been obtained exactly in Ref.~\onlinecite{igloi96,igtukasz97}

Assuming the same finite-size behavior for the first component of the 
eigenvectors
corresponding to the two lowest excitations, $\Phi_1(1)\sim\Phi_2(1)\sim
L^{-x_{m_s}}$,
the surface energy density exponent also follows
\begin{equation}
x_{e_s} =z+2x_{m_s}.\label{eq-44}
\end{equation}

The~correspondence~between~the~classical two-dimensional system and the quantum chain
shows that the correlation length in the Euclidean time direction (along the layers)
is given by the inverse of the smallest non-vanishing gap in the Hamiltonian 
spectrum, {\it i.e.} $\xi_\parallel=1/\varepsilon_1$ at $\lambda_c$. In the
low-temperature phase $\lambda>\lambda_c$, the lowest excitation $\varepsilon_1$
vanishes exponentially due to the asymptotic degeneracy of the ferromagnetic ground state in
the thermodynamic limit, and the parallel correlation length is
thus given by the next fermion excitation:
\begin{equation}
\xi_\parallel=\frac{1}{\varepsilon_2}.\label{eq-60}
\end{equation}
In the previous sections, we found the marginal behavior of the surface 
magnetization at the aperiodic critical point, and its crossover towards the usual
ordinary surface transition fixed point in the ordered phase.
We can expect here, as in the usual layered problem, a strong anisotropic 
scaling behaviour responsible for a special behavior of $\xi_\parallel$ at the
aperiodic fixed point. It can
be studied through the temperature dependence when $\lambda>\lambda_c$ ({\it via}
$\varepsilon_2$) or by finite-size scaling at the critical point $\lambda=\lambda_c$
({\it via}
$\varepsilon_1$).

\begin{figure}
\epsfxsize=8.6cm
\begin{center}
\mbox{\epsfbox{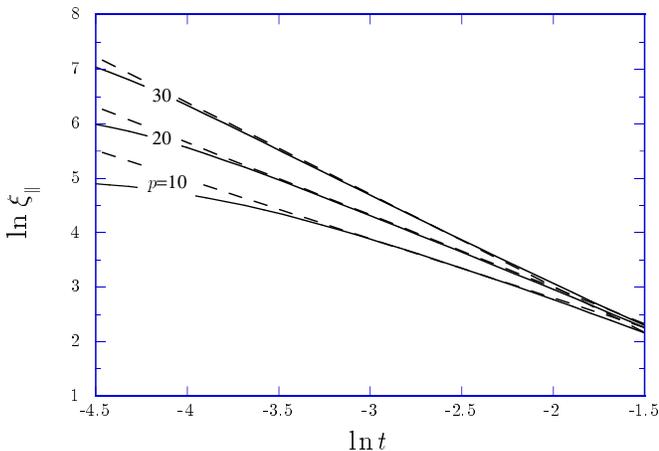}}
\end{center}
\caption{Temperature dependence of the parallel correlation length
 on a log-log scale  in the ordered phase
$\lambda>\lambda_c$. The deviation from the asymptotic linear behavior
(dashed lines) when
$t\to 0$ is a finite size effect.}\label{corrpar}  
\end{figure}

With free boundary conditions, the system (\ref{eq-3}) can be rewritten in
a  matrix form $(\mat{A}-\mat{B})(\mat{A}+\mat{B}){\boldmath\Phi}_\alpha=
\varepsilon_\alpha^2{\boldmath\Phi}_\alpha$. Here, the standard 
notation has been used.\cite{lieb61} Due to the tridiagonal structure of the 
``excitation matrix'' $(\mat{A}-\mat{B})(\mat{A}+\mat{B})$, the recursion relations for the components of ${\boldmath\Phi}_\alpha$
take the form:\cite{schmidt57}
\begin{eqnarray}
\pmatrix{{\mit\Phi}_\alpha (l)\cr
{\mit\Phi}_\alpha (l+1)\cr}&=&\mat{T}_l\pmatrix{{\mit\Phi}_\alpha (l-1)\cr
{\mit\Phi}_\alpha (l)\cr},\nonumber\\
&=&\pmatrix{0&1\cr
s_l&t_l (\varepsilon_\alpha )\cr}\pmatrix{{\mit\Phi}_\alpha (l-1)\cr
{\mit\Phi}_\alpha (l)\cr}
\label{eq-53}
\end{eqnarray}
with
\begin{equation}
s_l=-{\lambda_{l-1}\over\lambda_l},\quad t_l (\varepsilon_\alpha
)={\varepsilon_\alpha^2 -1-\lambda_{l-1}^2\over\lambda_l}\ ,\quad
2\leq l\leq L-1
\label{eq-54}
\end{equation}
and boundary terms 
\begin{eqnarray}
\pmatrix{{\mit\Phi}_\alpha (1)\cr
{\mit\Phi}_\alpha (2)\cr}&=&\mat{T}_1\pmatrix{{\mit\Phi}_\alpha (0)\cr
{\mit\Phi}_\alpha (1)\cr},\nonumber\\
\pmatrix{{\mit\Phi}_\alpha (L)\cr
\lambda_L{\mit\Phi}_\alpha (L+1)\cr}&=&\mat{T}_L\pmatrix{{\mit\Phi}_\alpha 
(L-1)\cr
{\mit\Phi}_\alpha (L)\cr},
\label{eq-55}
\end{eqnarray}
\begin{eqnarray}
{\mit\Phi}_\alpha(0)&=&0\nonumber\\
\lambda_L{\mit\Phi}_\alpha(L+1)&=&0.
\label{eq-56}
\end{eqnarray}
Finally, one has 
\begin{equation}
\pmatrix{{\mit\Phi}_\alpha (L)\cr
f(\varepsilon )\cr}=\mat{T}_L\dots \mat{T}_2 \mat{T}_1\pmatrix{0\cr 1\cr}=
\mat{M}_L\pmatrix{0\cr 1\cr}
\label{eq-57}
\end{equation}
where the eigenvectors are not normalized at this step and the excitation 
energies are deduced from the positive zeroes of $f(\varepsilon)$.
In a multilayered structure, within a given layer of size $p$, the same matrix appears $p-1$ times
and in the case $p=q$, Eq.~(\ref{eq-57}) becomes
\begin{eqnarray}
\mat{M}_L =&&\mat{T}_L (\mat{P}_{f_L})^{p-2}\mat{Q}_{f_L
,f_{L-1}}(\mat{P}_{f_{L-1}})^{p-1}\dots\nonumber\\
&&\dots\mat{Q}_{f_3 ,f_2}(\mat{P}_{f_2})^{p-1}\mat{Q}_{f_2 ,f_1}
(\mat{P}_{f_1})^{p-1}\mat{T}_1
\label{eq-61}
\end{eqnarray}
where $\mat{P}_{f_k}$ is a ``propagation'' matrix inside layer $k$ and 
$\mat{Q}_{f_{k+1} ,f_k}$ a ``transfer matrix'' from layer $k$ to layer $k+1$.
We can easily give a closed form for the propagation terms:
\begin{equation}
\mat{P}_{f_k}=\pmatrix{0&1\cr
-1&t_k (\varepsilon )\cr},
\label{eq-62}
\end{equation}
after diagonalization, leads to
\begin{equation}
(\mat{P}_{f_k})^{p-1}=\mat{S}_{f_k}\pmatrix{\alpha_+^{p-1}& 0\cr
0&\alpha_-^{p-1}\cr}\mat{S}^{-1}_{f_k}
\label{eq-63}
\end{equation}
where $\alpha_\pm$ are the eigenvalues of $\mat{P}_{f_k}$ and $\mat{S}_{f_k}$
the  changing of basis matrix.
\begin{figure}
\epsfxsize=8.6cm
\begin{center}
\mbox{\epsfbox{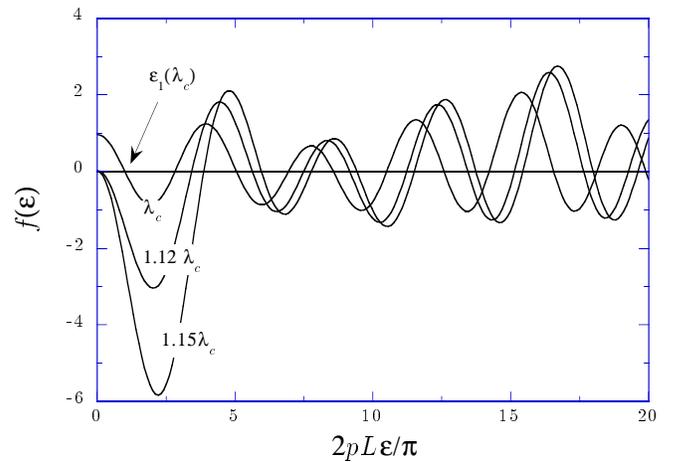}}
\end{center}
\caption{Typical shape of the function $f(\varepsilon)$ for period-doubling sequence,
with $L=2^4$, $p=q=2$, $r=1.1$, and several values of $\lambda$ in the
ordered phase.}\label{excit}  
\end{figure}
It is now possible to get $\mat{M}_L$ for any value of the layer sizes $p$, and, in the case
of large values of $p$, this technique is quite efficient compared to a direct diagonalization
of the excitation matrix. Some numerical troubles can nevertheless occur, since the function 
$f(\varepsilon)$ exhibits oscillations whose amplitude increases sharply with
the sizes $L$ and $p$ while the period decreases rapidly.
An exemple is shown in Fig.~\ref{excit} for a small size.
It is clear from Fig.~\ref{excit} that, as $\lambda$ increases from its critical value,
the lowest excitation $\varepsilon_1$ decreases while the next one, $\varepsilon_2$,
increases. The temperature behaviour of $\xi_\parallel$ is in agreement with
a modified power-law $\xi_\parallel\sim t^{-z}$, where the anisotropy exponent $z$
is assumed to be the sum
\begin{equation}
z(r,p,q)=\beta_s^{(m)}(r,p,q)+\beta_s^{(m)}(r^{-1},p,q)
\label{eq-64}
\end{equation}
as in the layered system. The temperature dependence of $\varepsilon_2^{-1}$
is shown on a log-log scale in  Fig.~\ref{corrpar} in the ordered phase
$\lambda>\lambda_c$. From the linear behavior, we can obtain the anisotropy exponent $z$.
The numerical results deduced from the slopes of the asymptotic linear behavior 
  are given in Tab.~\ref{tab1}. The numerical data are in agreement 
  with Eq.~(\ref{eq-64}).

\vbox{
\narrowtext
\begin{table}
\caption{Parallel correlation length exponent $\nu_\parallel$, deduced
from the slopes of log-log plots of $\xi_\parallel$ {\it v.s.} $t$,
 compared to the anisotropy exponent $z$ for the period-doubling sequence.
\label{tab1}}
\begin{tabular}{llll}
$r$ & $p=q$ & $\nu_\parallel$ & $z$ \\
\tableline
1.1&10&1.102&1.072\\
1.1&20&1.286&1.273\\
1.1&30&1.493&1.575\\
0.9&10&1.099&1.087\\
0.9&20&1.334&1.330\\
0.9&30&1.626&1.686\\
\end{tabular}
\end{table}
\narrowtext
 
A finite-size-scaling study provides an alternate numerical calculation of
these exponents.} It has been done in Tab.~\ref{tab2} for the anisotropy exponent,
but also for the surface 
magnetization and energy density exponents, in order to check the conjecture 
(\ref{eq-44}) for the multilayered system. The layer widths have been limited 
to small
values in order to reach large enough values of $L$ ($2^{10}$ for the period 
doubling sequence
for example) and extrapolation to infinite size has been performed, using the
BST algorithm, which is  efficient for finite-size-scaling analyses.~\cite{henkelschutz88} 

\vbox{
\narrowtext
\begin{table}
\caption{Anisotropy, surface  magnetization and surface energy density exponents 
 deduced
from a finite-size-scaling study for the period-doubling sequence.
\label{tab2}}
\begin{tabular}{llllllll}
&&\multicolumn{2}{c}{$z$}&\multicolumn{2}{c}{$x_{m_s}^{(m)}$}&\multicolumn{2}{c}{$x_{e_s}^{(m)}$}\\
$r$ & $p=q$ & num. & theor. & num. & theor. & num. & theor.\\
\tableline
2&2&1.15(2)&1.149&.58(3)&.574&2.29(1)&2.298\\
2&3&1.33(2)&1.322&.69(5)&.661&2.62(3)&2.644\\
3&2&1.36(2)&1.357&.70(4)&.678&2.70(2)&2.713\\
3&3&1.75(2)&1.737&.90(4)&.869&3.45(2)&3.474\\
\end{tabular}
\end{table}
\narrowtext

\section{Conclusion}
\label{sec:ccl}
We have presented the study of the influence of layer widths $p$ and $q$ in aperiodic
multilayered Ising models. The perturbations under consideration correspond to
vanishing wandering exponents, and thus lead to marginal critical behaviors in
the two-dimensional Ising model. From a finite-size scaling analysis of the aperiodic
series which defines the surface magnetization, we obtained the corresponding
critical exponents which continuously vary with the modulation ratio. This 
defines the {\it aperiodic fixed point behavior} characterized by an anisotropic
scaling which has been studied numerically.

When the system, in the ordered phase, is moved away from this critical point, 
the surface magnetization exhibits a crossover towards another regime, governed
by the ordinary surface transition fixed point. We have shown that the crossover
is well described by scaling functions of a single scaled variable which
obey well-defined asymptotic behaviors.

Aperiodic multilayers are now feasible by molecular beam epitaxy,~\cite{maj91}
and although nothing experimental has been done up to now to study the critical
behavior of such systems,
 crossover effects between the two regimes should be experimentally observable.

One should finally mention that the same type of study for relevant
aperiodic modulations would be interesting especially when an enhancement of
the modified couplings
 close to the surface leads to first order surface transitions in a 
 two-dimensional
 system. This will be the subject of a further investigation.
}

\acknowledgments We  are indebted to Lo\"\i c Turban and Dragi Karevski
 for valuable discussions.



\begin{references}
\bibitem{notet} The deviation $t$ from the critical point is 
defined as a positive quantity in the ordered phase.
\bibitem{z} This relation states that $z$ is given by the sum of the 
scaling dimensions associated to the left ($x_{m_s}(\lambda_c)$) and
the right surfaces ($\bar x_{m_s}(\lambda_c)=x_{m_s}(\lambda_c^{-1})$), the latter one being deduced
from the former by the transformation $r\to r^{-1}$, which is equivalent to
$\lambda_c\to\lambda_c^{-1}$.
\end{references}
\end{document}